\newcommand{\draftnote}{}

\documentclass{ws-ijmpa}
\usepackage[super,compress]{cite}
\usepackage{graphicx}
\usepackage{xcolor}
\usepackage{url}
\usepackage{wrapfig}
\usepackage{hyperref}

\begin{document}
\markboth{Szumila-Vance, Dutta, Miller, Sargsian}{Seeing Through the Nucleus}

%
\catchline{}{}{}{}{}
%

\title{Seeing Through the Nucleus:\\ A Review of Color Transparency Phenomena}

\author{Holly Szumila-Vance}
\address{Department of Physics, Florida International University, 11200 SW 8th St\\
Miami, FL 33199, USA\\
hszumila@fiu.edu}

\author{Dipangkar Dutta}
\address{Department of Physics and Astronomy, Mississippi State University, 355 Lee Boulevard\\ 
Mississippi State, MS 39762, USA\\
dd285@msstate.edu}

\author{Gerald A. Miller}
\address{Department of Physics, University of Washington, 1410 NE Campus Pkwy\\
Seattle, WA 98195-1560, USA\\
miller@uw.edu}

\author{Misak Sargsian}
\address{Department of Physics, Florida International University, 11200 SW 8th St\\
Miami, FL 33199, USA\\
sargsian@fiu.edu}

\maketitle

\begin{history}
\end{history}

\begin{abstract}

We review the current status of the phenomenon of Color Transparency (CT), a fundamental consequence of the description of hadrons from Quantum Chromo Dynamics.  
CT refers to the vanishing of final (and/or initial) state interactions with the nuclear medium for exclusive process at sufficiently high enough momentum transfers. We discuss the current experimental observations relating to CT and their theoretical implications for other high energy processes. Future CT experiments and facilities are also described.   

\end{abstract}



\section{Introduction}
Color transparency (CT) is an interesting and surprising prediction of Quantum Chromodynamics (QCD), the fundamental theory of the strong interaction,  related to deep questions regarding hadronic structure. 
Color transparency has a rather unusual name. One might think that this is about objects that have color and are transparent, but it is really about how objects without color are transparent. More technically, color transparency refers to the phenomenon in which hadrons, produced or struck in coherent,  high-momentum processes, can briefly fluctuate to a reduced transverse size, color-neutral configuration that only barely interacts with the nuclear medium. In this scenario, the effective cross section of the hadron is reduced, leading to an increased probability of escape from the nucleus without further interaction. The existence of color transparency relies on the hypothesis that a hadronic wave function contains components that are of much smaller spatial extent  than the average size, and that those components play an essential role in producing coherent, high-momentum transfer processes.
 
Over the past several decades, numerous experimental efforts have aimed to observe and quantify CT effects using electron- and hadron-induced reactions. These include measurements of nuclear transparency in quasielastic $(e,e'p)$ reactions, meson electroproduction, photoproduction processes, and exclusive hadron scattering. While early results from the meson electroproduction experiments showed hints for evidence of CT, ongoing and future experiments are needed to confirm the observation and explore the characteristics of CT. One experiment using a high-energy pion beam observed a strong signal consistent with CT predictions, see Section~\ref{sect:pidiffdiss}. CT has received a revived interest in recent years after the null observation of CT in quasielastic electron-induced proton knockout~\cite{PhysRevLett.126.082301} at Jefferson Lab. 

This review aims to provide an overview of the current status of CT studies. We begin by outlining the theoretical foundations and the expected signatures of CT across different reaction channels. We then examine the experimental evidence gathered to date, highlighting recent progress, key findings, and open questions. Special emphasis is given to new approaches and strategies including a discussion of different facilities which promise to deepen our understanding of this fundamental QCD phenomenon. Through this blend of experimental and theoretical developments, we seek to clarify the role of CT and the observation of small-sized states in the hadronic wave function.

\section{What is Color Transparency and Why is it Interesting? }	
The CT phenomenon is one of the most interesting properties of QCD. It is not a simple extension of charge transparency, as observed in Quantum Electrodynamics (QED)\cite{Perkins:1955}. This is because of non-Abelian nature of the strong interaction, which is governed by the  color $SU(3)$ gauge theory. 

We provide a brief explanation here. The strong interaction between hadrons and nuclei generally leads to shadowing effects that reduce the ability for a hadron to transverse a nucleus without losing energy or being absorbed. However, in the special situation of high-momentum-transfer coherent processes these interactions can be turned off, causing the shadowing to disappear and the nucleus to become quantum mechanically transparent. This phenomenon is known as color transparency. In more technical language, CT is the vanishing of initial- and final-state interactions, predicted by QCD to occur in high-momentum-transfer quasielastic reactions. These are coherent interactions in which one adds different contributions to obtain a total scattering amplitude. Under these specific conditions  the effects of gluons emitted by small-sized color-singlet systems vanish because of color neutrality.
For baryons, this neutrality is different than the charge cancellation of QED, because of the color $SU(3)$ property of the theory. The name ``color transparency" is rather unusual. One might think that it is about transparent systems that have no color, but it is really about how a medium can be transparent to objects without color.

There are three necessary conditions for CT to occur~\cite{Jennings:1989hc,Jennings:1990ma}: (1) High momentum transfer reactions occur via components of hadronic wave functions that are small-sized. (2) Small-sized objects have small cross sections. (3) Small-sized objects are not eigenstates of the Hamiltonian, and so must evolve with time. Since they start out small, their size must grow. This expansion must be a small effect for CT to occur. These different conditions are explained below.

\subsection{Color Neutrality in QCD}
The concept of strongly interacting elementary particles involves the existence of hidden color charges which form exact $SU(3)$ color gauge symmetry. This symmetry predicts the existence of two types of singlets which are combinations of (color)-(anti-color) as realized in mesons and three color (i.e.RGB) configurations as realized in baryons. 
Theoretically, one can still have colored mesons and baryons (such as hidden color components in two-baryon systems), but for not  fully understood dynamics, related perhaps to 
``peculiar infrared properties of non-Abelian gauge theories"\cite{Weinberg:2004kv} nature only permits  colorless mesons and baryons in the free state.

Since local $SU(3)$ color gauge theory is the origin of strong interaction as mediated by gluons, one expects that the strength of the interaction is proportional to the volume occupied by the gluonic field in the hadron. Phenomenologically, this expectation is supported by the fact that light quark mesons (e.g. $\pi$, $\rho$ $\omega$-mesons) have approximately the same total hadronic scattering cross sections and approximately the same sizes. The same is apparently true for baryons~\cite{PhysRevLett.58.1612,POVH1990653}. This supports the picture in which the strength of the strong interaction is proportional to the sizes of the interacting hadrons.

It is worth noting that a similar picture exists for electromagnetic interactions. For example, the electric potential caused by  a dipole is $\propto {d\over r^3}$ where $d$ is the separation between two oppositely charged point particles, and $r$ is the distance from the dipole system. For mesons, the cancellation of the effects of red and anti-red quarks is similar to the charge cancellation of QED. The situation for baryons is more interesting because the cancellation would arise from quarks of three different, non-zero color. This makes the hunt for color transparency reactions that  involve protons very interesting. The color cancellation of $SU(3)$ is perhaps the only property of QCD that has not been verified.

\subsection{Point Like Configurations (PLC) in Hadrons  and the Minimal-Fock Component Description of PLC}
A common property of hadrons is that they all have valence quarks that define their quantum numbers and contain potentially unrestricted additional sea quarks and gluons. This enables the possibility that there exists a hadronic component consisting of only valence quarks at very short separations that are color neutral and carry the quantum numbers of the hadron.  We will refer to such configurations as Point Like Configurations (PLCs). The PLC component is meant to represent the smallest-sized component of possible configurations of the free hadron, since substantial contributions of sea quarks and gluons are necessary to make up the finite size of the hadron.

Early lattice calculations provide evidence  for the existence of hadronic PLCs~\cite{IOFFE1981317,FUCITO1982407}. This is because  initial configurations were chosen to be PLCs.  For protons all three quarks were taken to be at the same space-time location. For mesons an initial quark-anti-quark PLC was chosen. These configurations evolve in Euclidean time to eventually  become the physical system, showing that PLCs are not orthogonal to the physical wave function. Later, finite-size configurations were used to accelerate that evolution\cite{Gusken:1989qx}. 

\begin{figure}[htb]
\begin{center}
\includegraphics[width=1.8in]{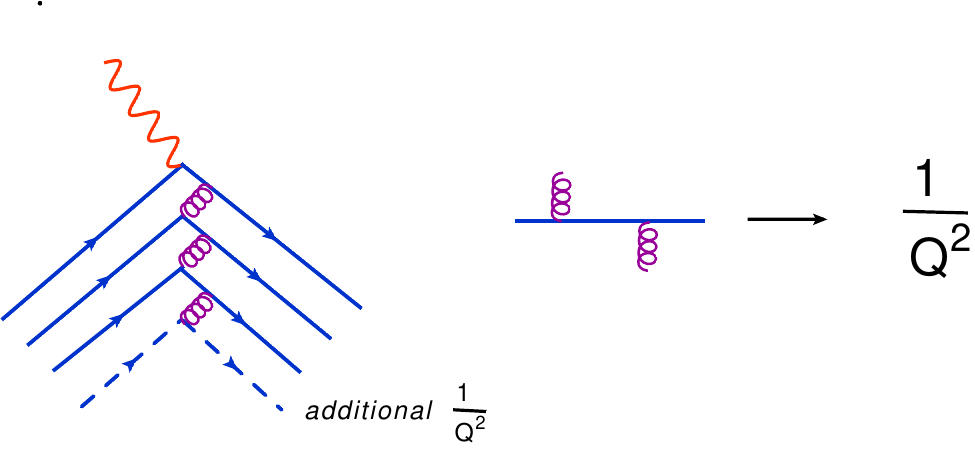}
\hspace{1.4cm}
\includegraphics[width=1.8in]{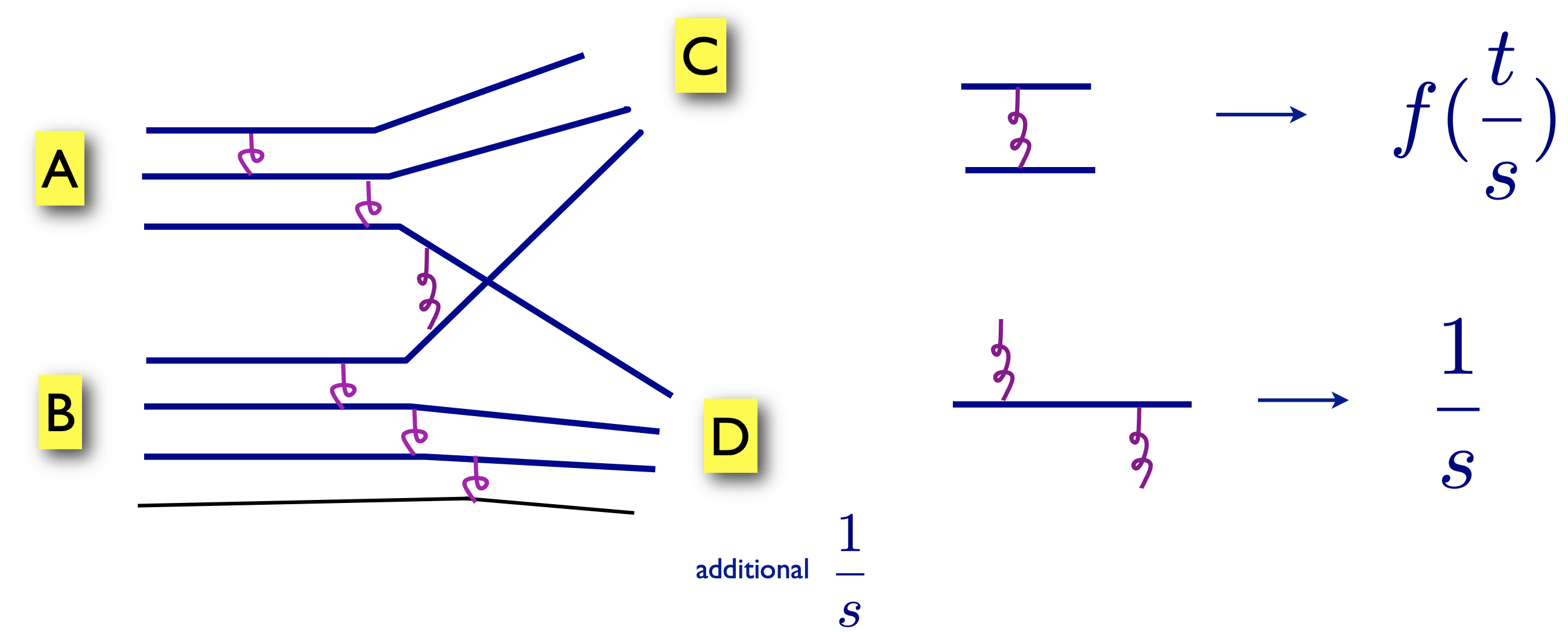}
\caption[]{\label{fig:mfc} Left: Exclusive electroproduction reaction at large $Q^2$.
Right: Hard exclusive hadron scattering at large $s$ and $t$.}
\end{center}
\end{figure}

One of the unique properties of hard exclusive processes, in which the final state of the process is constrained by the mass of the produced hadrons, is that if the scattering process is dominated by the interaction of an external probe with the valence quarks in the hadron,  then  the hadron component with the minimal number of quarks has the largest contribution to the process. This dominance exists both for exclusive electroproduction processes at large $Q^2$ (Fig.\ref{fig:mfc},left panel) and hard hadronic exclusive processes at large 
$s$ for fixed center of mass scattering angles (see Fig.\ref{fig:mfc}, right panel).

As shown in Fig.\ref{fig:mfc} (left panel), in a perturbative QCD analysis, each additional quark line appearing in the interaction process introduces an additional factor of ${1\over Q^2}$ that suppresses  the amplitude. The same is true for hard exclusive $AB\to CD$ hadronic reactions 
(Fig.\ref{fig:mfc}, right panel), in which case the additional quark-line involved in the hard scattering introduces an additional ${1\over s}$-suppression. Thus, the dominance of the PLC in  hard interactions is a natural property of perturbative QCD.
The tendency for  various wave function models to have a PLC was assessed in~\cite{Frankfurt:1993es}. The validity of pQCD is not a necessary requirement for the existence of a PLC. Nevertheless, we use pQCD for a couple of paragraphs.

The next question is whether the minimal Fock component represents a small sized component of the hadron involved in the scattering - commonly referred to as a Point-Like Configuration (PLC). A conjecture that the minimal Fock-state component involved in the hard scattering is a PLC follows from the fact that the quark that absorbs the high momentum from the probe will accelerate, thus radiating gluons. To exclude these gluons in the final state of the reaction (since the considered process is exclusive), one needs  nearby quarks at distances $\sim {1\over Q}$ to  absorb  the radiated gluons. Therefore, one expects that the minimal Fock component should be of size $\sim {1\over Q}$. 

In the hard processes under consideration, the probe interacts with individual valence quarks in the hadron and transfers a momentum $\sim \sqrt{Q^2}$. With a swift change of its original momentum in the hadron, the interacting quark will radiate gluons. Considering only exclusive channels, these gluons should be absorbed by other quarks to allow the specific final hadronic state.  

\begin{figure}[htb]
\begin{center}
\includegraphics[width=1.8in]{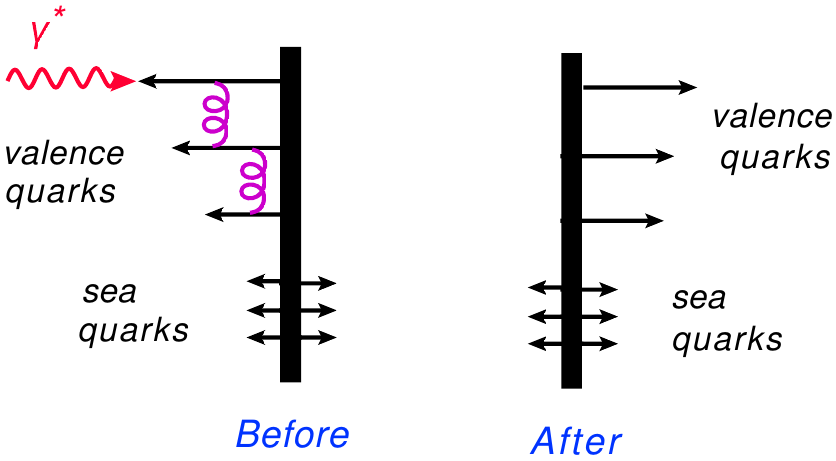}
\hspace{1.8cm}
\includegraphics[width=1.8in]{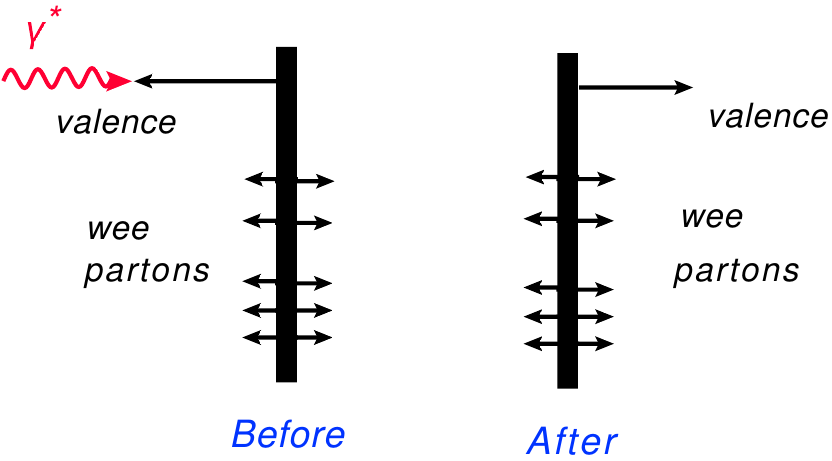}
\caption[]{\label{fig:mfc_feynman}(left panel) Minimal Fock Component mechanism of hard scattering.(right panel) Feynman Mechanism of of hard scattering.}
\end{center}
\end{figure}

Dimensional analysis indicates that such gluons should have momenta of the order of $Q$. In order to suppress the radiative gluons in the final state of the reaction, the other valence quarks should be at close proximity (at the distances of ${1\over Q}$) to absorb the radiated gluons. Within the Minimal Fock-state component of hard scattering, this results in a selection of three valence quarks at relative distances of ${1\over Q}$ naturally forming a PLC. Such a process corresponds to the left panel of Fig.\ref{fig:mfc_feynman}. 

Determining the existence of a PLC and the closely related mechanism responsible for high-momentum transfer exclusive reactions are the main goals of color transparency research. If the radiating gluonic field is not dominated by hard gluon components, then the field can be absorbed by the wee partons as shown in Fig.\ref{fig:mfc_feynman}, right panel.  This corresponds to the Feynman mechanism of electroproduction in which the virtual photon is absorbed by a leading quark in the hadron without disturbing the distribution of wee partons that exist in both positive and negative momentum fraction space. As it follows from the figure, such a mechanism does not produce a PLC, and no CT effects are expected in quasielastic processes in nuclei if the underlying reaction is dominated by the Feynman mechanism.

An analogous mechanism exists for hard hadron-hadron interactions, where instead of the minimal-Fock state component mechanism of the scattering (Fig.\ref{fig:mfc}), the hard scattering proceeds through the individual interaction of quarks from each hadron. Such a mechanism is referred to as the Landshoff mechanism and does not require a PLC formation. Thus, no CT effects will be observed if hadronic scattering takes place in the nuclear medium. Such a mechanism could explain the observed decrease in transparency with increasing proton beam energy as observed in the Brookhaven $(p,pp)$ experiments~\cite{Leksanov:2001ui,Aclander:2004zm,Jain.Pire.Ralston.2022}.

The theoretical approach used for quantitative investigations of the effects of the re-absorption of radiated gluons in hard scattering  was introduced in studies of electromagnetic processes to account for  photon radiation. It was shown in  Ref.~\cite{Botts:1989kf,Botts:1989nd,Li:1992nu} that reabsorption of the radiated field can be evaluated through the Sudakov form factors that account for 
the relative proportion of hard and gluon radiative fields in the scattering process.

\subsection{Expansion of the PLC}
\label{PLCexpansion}
The PLC is not an eigenstate of the QCD Hamiltonian and necessarily changes with time. Assuming  that the system starts out as small-sized, any change must be an expansion in size. If energies are not significantly large, then the expansion will take place during the PLC's propagation in the nuclear medium. Quantitatively, it will reveal itself as a measurable rescattering of the expanding PLC in the nuclear medium as compared to the contribution in which no final state 
interaction (FSI) occurs (usually referred to as the Plane Wave Impulse Approximation (PWIA)).

Considering the FSI of a multi-GeV PLC in the nuclear medium, the rescattering is proportional to the total cross section of the PLC-nucleon scattering interaction which is defined mainly by the square of the PLC's transverse size. In the quantum diffusion model~\cite{Farrar:1988me}, it is expected that the square of the transverse size of the PLC is approximately proportional to the distance $z$ that the PLC traveled from the point of the hard interaction:
\begin{equation}
    \sigma^{PLC}(z) = \left(\sigma^{hard} - {z\over l_c}(\sigma - \sigma_{hard})\right) \Theta(l_c - z) + 
    \sigma\Theta(z-l_c),
    \label{QDM}
\end{equation}
where $\sigma^{PLC}(z)$ is the total cross section of the fully formed hadron scattering from the  nucleon, and $\Theta()$ functions 
are introduced to provide a smooth transition from 
the PLC to the soft regime of rescattering. Here, 
$l_c$ is the coherence length that characterizes the phase at which PLC holds its identity and depends on the nature of the hard scattering. 
Intuitively, $l_c$ is the distance over which the hard scattering can produce inelastic states with mass $M_X$ without disturbing the nuclear environment. It can be estimated through the propagator between the hard scattering and the first rescattering point as
\begin{equation}\label{lc}
l_c\sim {2p_{lab}\over M_X^2 - M_h^2} = {2p_{lab}\over \Delta M^2},
\end{equation}
where in the right part of the equation we introduce a largely unknown  parameter $\Delta M^2$ that characterizes the difference of mass squares between the ground state of the hadron and the average masses in the intermediate state that are
large enough to ``liberate quarks" into forming a PLC. It is worth noting that the initial estimates of $\Delta M^2 = 0.7-1 $GeV$^2$ were based on the additive quark model. The slope of the 
Regge trajectory is ruled out by recent experiments on $(e,e^\prime p)$ scattering\cite{PhysRevLett.126.082301} indicating a significantly larger value of $\Delta M^2$. 
The physical property of the $\Delta M^2$ parameter is that the excited baryonic masses in the intermediate state are large enough that their coherent superposition results in a liberation of quarks in the PLC due to saturation of the 
baryonic mass spectrum. However, the role of the 
quantum numbers in such a saturation is not 
well understood. For example, if one requires only a spectrum of positive parity baryons with half spin and isospin, then the spectrum consists of N(1440) and N(1910) and results in a 
$\Delta M^2 \gtrsim 3~GeV^2$, which is more consistent with the recent $(e,e^\prime p)$ experiment\cite{PhysRevLett.126.082301}.

The complementary approach to account for the 
coherence effect and the handling of the expansion is to use a hadronic 
basis\cite{Jennings:1989hc,Jennings:1990ma,Frankfurt:1994kt}. Following the completeness postulate, the PLC  can be expressed as an expansion in terms of  an orthonormal basis of eigenstates. 
For the baryonic sector, such eigenstates correspond to the baryons that comprise the spectrum of the ground and excited states with the same spin and isospin. This includes  meson-baryon eigenstates. In the case of a baryonic PLC, one can express the quantum state of a PLC as:
\begin{equation}
\mid PLC\rangle_z = \sum\limits_{i=0}^{N}\alpha_i(z)\mid N_i\rangle
\label{hbasis}
\end{equation}
where $\mid N_i\rangle$ corresponds to the ground state (nucleon) and the excited state resonances. The sum is also meant to include the continuum. The condition that the above configuration is a PLC is that the initial ($z=0$) matrix element of the scattering amplitude of a PLC with any baryon (or hadron) vanish, i.e.
\begin{equation}
\langle N_i\mid \hat T \mid PLC\rangle_{z=0} = 0
\label{ressum}
\end{equation}
The coefficients, $\alpha_i(z)$, of the expansion in Eq.~(\ref{hbasis}) contain the phase factors that account for the mass differences that appear in the masses of baryonic resonances.
\begin{equation}\label{az}
a_i(z) = a_i \exp{\left(-i{m_i^2-m_0^2\over 2 P}z\right)}
\end{equation}
The advantage of such a formulation of the time-dependence of the PLC  is that it  naturally describes the expansion of the PLC at finite energies as it propagates through the nuclear medium. Note that Eq.~\eqref{az} accounts for the coherence length of Eq.~\eqref{lc}.

Several models of the expansion were proposed. The start of a more fundamental approach was studied in~\cite{MAKOVOZ1995622} using $SU(2)$ lattice gauge theory with Wilson fermions in the quenched approximation. The wave packet was modeled by a point hadronic source. The procedure is to determine the Euclidean time ($t$), pion channel, and Bethe-Salpeter amplitude and then evaluate the effects of a soft interaction of a small-sized wave packet with a pion. A superposition of three states was found sufficient to reproduce a reduced-size wave packet. Using this superposition allowed an  analytic continuation to  real time. The matrix elements of the soft interaction between the excited and ground states was found to decrease rapidly with the energy of the excited state, a result favoring the existence of color transparency. The use of modern lattice QCD techniques to study the existence and expansion of the postulated PLC would be greatly appreciated.  

Initial attempts to model PLCs through resonances were made in Refs.~\cite{Jennings:1989hc,Jennings:1990ma,Frankfurt:1992zp} where it was assumed 
that the finite number of baryonic resonances will saturate the sum in Eq.~(\ref{ressum}) and that the closure approximation is applicable. It is worth mentioning that the three-state model~\cite{Frankfurt:1992zp} was perhaps the most successful since it satisfied the exact condition that one of the matrix elements of $\hat T$ is exactly zero. A realistic baryon spectrum that included an infinite number of states was used to study the expansion phenomenon in ~\cite{Jennings:1992hs}. 
The results were similar to those of the quantum diffusion model~\cite{Farrar:1988me}, predicting effectively $\Delta M^2 \gtrsim 3$~GeV$^2$.  In  more recent work\cite{Caplow-Munro:2021xwi} the study of the expansion aspect of PLC
using superconformal baryon-meson symmetry and light-front holographic QCD resulted in $\Delta M^2 \approx 3.5$~GeV$^2$. It is worth noting that such values of $\Delta M^2$ are consistent with the non-observation of CT phenomena in $A(e,e^\prime p)X$ experiment for values of $Q^2$
up to about $14-15$~GeV$^2$\cite{PhysRevLett.126.082301}.

\section{Color transparency in various processes}
Here we summarize the experimental efforts for exploring CT in different processes. While some experiments are detailed in the previous review~\cite{DUTTA20131}, we highlight here the processes that are relevant for extending this discussion in light of recent results. 

\subsection{The one that worked- coherent production of two jets by high-energy pions}\label{sect:pidiffdiss}
The process $\pi A \to A+ 2~{\mbox{jets}}$, in which the nucleus remains in its ground state provides an example of the ability to observe color transparency when all of the requirements mentioned previously are satisfied.
The pion as a $q\bar q $ state has a larger probability to form a PLC than the proton.  At high energies, the effects of expansion can be ignored. The ability to detect two jets as a function of  high relative momentum allows the experimenter to ensure that a PLC was indeed formed. As a result of these favorable conditions, the Fermilab experiment E791 ~\cite{E791:2000kym,Aitala:2000hb} found very strong signals for color transparency- a very large ratio of the Pt to C cross sections that provided the necessary signature \cite{Frankfurt:1993it} of the effect. For further details, see the review~\cite{DUTTA20131}. Since the experiment was carried out at very high energies it did not provide any information about the onset of CT. The onset of CT has been the focus of all the recent experiments.

\subsection{Recent experiments}
Measurements in the intermediate energy regime are of direct interest to the observation of the onset of CT. The most comprehensive review~\cite{DUTTA20131} of the experimental efforts for exploring CT was covered over a decade ago. Since that time the only new, published experimental result was from a recent Jefferson Lab Hall C experiment which ruled out the observation of CT effects in quasielastic proton knockout reactions on a carbon target up to a $Q^2<14$~GeV$^2$~\cite{PhysRevLett.126.082301,PhysRevC.108.025203}. The null observation was consistent with the conventional nuclear physics descriptions excluding CT effects and was somewhat surprising. 

\subsubsection{Direct searches for evidence of CT in protons}
The kinematics of the recent Hall C electron-scattering proton knockout experiment were chosen to overlap with the proton momenta in the BNL experiments that used wide angle $(p,pp)$ scattering. The BNL $(p,pp)$ experiments observed an initial rise in the transparency for an effective proton momentum of $p_p= 6 - 9.5$~GeV/$c$ that was consistent with the selection of a small size configuration and its subsequent expansion over distances comparable to the nuclear radius (also this expansion was consistent with that of the meson predictions)~\cite{Mardor:1998zf,Leksanov:2001ui,Aclander:2004zm}. However, the transparency then decreased with further increasing effective beam momentum and was not consistent with CT theory alone. The decrease in the transparency at higher momentum was reasonably, although not conclusively~\cite{Jennings:1993hw}, explained as an energy dependence of the free cross section~\cite{ralpire,JAIN199667,physics4020038} or as a possible resonance or threshold for a new scale of physics~\cite{brodsky_ch}. In any case, an added complication of the $(p,pp)$ measurement is that the incoming proton also suffers a reduction of flux that must be accounted for in extracting transparency effects. Thus, the electron beam proton knockout experiments at Jefferson Lab provided an ideal setup to further explore the onset of CT in protons because only the outgoing proton needs to be considered.

In the plane wave impulse approximation (PWIA) in quasielastic electron scattering, the proton is ejected without final state interactions with the residual $A-1$ nucleons. The measured $A(e,e'p)$ cross section would be reduced compared to the PWIA prediction in the presence of final state interactions, where the proton can scatter both elastically and inelastically from the surrounding nucleons as it exits the nucleus. The deviation from the simple PWIA expectation is used as a measure of the nuclear transparency. While additional effects such as nucleons in short-range correlations and the density dependence of the $NN$ cross-section will affect the absolute magnitude of the nuclear transparency, they have little influence on the $Q^2$ dependence of the transparency. The $(e,e'p)$ reaction is simpler to understand than the $(p,pp)$ reaction and immediately spurred a series of experiments first at SLAC ~\cite{Makins:1994mm,ONeill:1994znv} and then Jefferson Lab~\cite{Abbott:1997bc,Garrow:2001di} for a range of light and heavy nuclei. Further details of these experiments can be referenced in the previous review~\cite{DUTTA20131}.

In response to the observed lack of CT in protons, a workshop~\cite{ctworkshop} brought together the international community to interpret the results and re-evaluate CT predictions. The Feynman Mechanism, discussed further in Sect.~\ref{endpoint}, is one of the leading explanations of these results in their respective kinematics. 

While it is possible that at higher values of momentum transfer the dominant mechanism might be different than the Feynman mechanism~\cite{Brodsky:2022bum}, the conclusions stated in Ref.~\cite{Brodsky:2022bum} are flawed because the Eq.(12) of that reference is not correct.
One may compute the effective size transverse size $b_T^2(Q^2)$ as a function of momentum transfer~\cite{Frankfurt:1993es,PhysRevD.74.034015}.
This is given by 
\begin{equation}
    b^2_T(Q^2)=\frac{\langle H(Q^2)|\widehat b_T^2\,T_H(Q^2)|H\rangle}{\langle H(Q^2|T_H(Q^2)|H\rangle}
\end{equation}
where an initial hadron $H$ acquires a momentum transfer $q^2=-Q^2$ due to the action of a hard scattering operator, $T_H(Q^2)$, and the denominator is the hadronic form factor.
The error in Eq(12) of  Ref.~\cite{Brodsky:2022bum}
is that the operator $\widehat b_T^2$ is replaced by $-{d\over dQ^2}$ taken outside the overlap integral needed to compute the matrix element. 

\subsubsection{CT in meson electroproduction experiments}
The onset of CT is favored to be observed at lower energy in mesons than baryons since only two quarks must come close together, and the quark-antiquark pair is more likely to form a PLC~\cite{blattel93}. Moreover, the effects of PLC expansion are less significant than for protons~\cite{Caplow-Munro:2021xwi}. Pion electroproduction measurements at Jefferson Lab in the 6~GeV-beam era reported evidence for the onset of CT \cite{Clasie:2007aa} in the process $e + A \rightarrow e + \pi^+ + A^{*}$. The results of the pion electroproduction experiment showed that both the energy and the $A$-dependence of  nuclear transparency deviate from conventional nuclear physics and are consistent with models that include CT. The pion results indicate that the energy scale for the onset of CT in mesons is $\sim 1$~GeV. 

Furthermore, a Hall B CLAS experiment studied $\rho^0$-meson production from nuclei, and the results also indicated an early onset of CT in mesons~\cite{ElFassi:2012nr}. The transparency for incoherent exclusive $\rho^0$ electroproduction in carbon and iron relative to deuterium~\cite{ElFassi:2012nr} using a 5~GeV electron beam indicated an increase of the transparency with $Q^2$ for both nuclei. The rise in transparency for the $\rho^0$ was found to be consistent with predictions of CT by models~\cite{Frankfurt:2008pz,Gallmeister:2010wn} that had accounted for the increase in transparency for pion electroproduction. This Hall B experiment recently completed new data-taking at Jefferson Lab using a higher electron beam energy to extend the range of $Q^2$ for the $\rho^0$ transparency study. A discussion of the anticipated results for both the $\rho^0$ and $\pi^+$ are described in Section~\ref{sect:future}.

\subsubsection{CT in photoproduction reactions}
Ref. ~\cite{Larionov:2016mim}  demonstrated that studies of  semi-exclusive large angle photon - nucleus reactions: $\gamma + A\to h_1+h_2 +(A-1)^*$   with tagged photon beams of energies of 6-10~GeV  would probe several aspects of  QCD dynamics such as establishing the  range  of $t$ in which transition from soft to hard dynamics occurs, comparing the strength of the interaction of various mesons and baryons with nucleons at the energies of few GeV, and to directly look for CT effects. Such experiments are accessible by the Jefferson Lab energies.

A Jefferson Lab Hall A experiment explored this effect by using the electron beam on a copper radiator to generate an untagged beam of photons incident on $^4$He and $^2$H targets~\cite{Dutta:2003mk,DUTTA20131}. The transparency of the reaction of $\gamma n\rightarrow\pi^- p$ was taken as a ratio of the $^4$He/$^2$H targets and studied as a function of the $|t|$-dependence at $|t|<2.5$~GeV$^2$ for fixed center-of-mass scattering angles of 70$^\circ$ and 90$^\circ$. The transparency dependence with $|t|$ was compared to Glauber calculations and hinted at a deviation with increasing $|t|$. However, the statistical uncertainty was too large to draw a precise conclusion from the data. More recently, an experiment using the $\sim$8.5~GeV photon beam in Hall D at Jefferson Lab was able to explore photoproduction reactions for different processes in $^2$H, $^4$He and $^{12}$C. This experiment is discussed in further detail in Section~\ref{sect:future}.

\subsection{Feynman Mechanism {\it vs.} Color Transparency} \label{endpoint}
The striking experimental finding~\cite{PhysRevLett.126.082301} that color transparency does not occur in the $(e,e',p)$ reaction with momentum transfer up to 14.2~GeV$^2$  demanded an interpretation and evaluation of the consequences. Ref.~\cite{Caplow-Munro:2021xwi} aimed to provide such. The failure to observe color transparency could have arisen from two possibilities: (1) a PLC  was formed, but the expansion process caused final state interactions to occur before the outgoing proton could escape the nucleus, or (2) no PLC was formed. Ref.~\cite{Caplow-Munro:2021xwi}  studied the expansion aspect of CT using  a new formalism involving superconformal baryon-meson symmetry and light-front holographic QCD~\cite{Brodsky:2014yha}. This approach provides a complete spectrum of hadronic states and their light-front wave functions.
The sum over complete states indicated in Eq.~\eqref{hbasis} was achieved in closed form by using the Feynman path integral formulation. Calculations showed that  the expansion effects would not be large enough to cause significant final state interactions to occur. Therefore, it was concluded that a PLC was not formed in the experiment and that the Feynman mechanism involving the virtual photon absorption on a single high momentum quark is responsible for high momentum electromagnetic form factor of the proton.

\subsection{Color Transparency and Quark Gluon Plasma}
One of the important features of CT is the coherence length, $l_c$ which characterizes the phase at which PLC holds its identity. It represents a distance over which the initial hard scattering produces a sequence of inelastic hadronic states that interfere coherently without disturbing the nuclear environment. Thus, CT scans the range of hadronic masses which are in a coherent superposition, and the length of such coherence is inverse proportional (Eq.\ref{lc})  to the parameter $\Delta M^2 \approx M_X^2 - M_h^2$, where $M_h$-is the ground state mass of the hadronic spectrum. Here $M_X$ characterizes the limiting mass in the hadronic spectrum that is large enough to allow the   replacement of the coherent sum of hadronic states by 
a PLC.   

If CT is observed, $\Delta M^2$ may be evaluated within the quantum diffusion model (Eq.(\ref{QDM})) by extracting the value of $M_X$. Qualitatively, since $M_X$ represents the upper limit of the masses for the discrete hadronic spectrum, the discussion here pertains to how the knowledge of this mass helps to identify the transition  point from the hadronic to the quark-gluon phase.

Our discussion will be confined within Hagedorn's statistical bootstrap model\cite{Hagedorn:1965st,Hagedorn:1967tlw} according to which, the discrete hadronic spectrum is followed 
by an exponentially increasing spectrum of hadronic states. The consequence of this is that such a hadronic system has a limiting  temperature $T_C$ above which hadronic matter can not exist. Consideration of hadronic models which account for the 
internal quark-gluon structure (e.g. Ref.\cite{Cabibbo:1975ig}) indicate that this critical  temperature corresponds to the transition into the phase in  which  quarks are not confined.

The limiting mass $M_X$, that CT studies potentially can evaluate, corresponds to the threshold for the emergence  of the hadronic mass spectrum distribution, $\rho(m)$ satisfying 
the bootstrap condition, i.e.:
\begin{equation}
    \rho(m)\mid_{m>M_X}\sim  {const\over m ^{a}} e^{m/kT_0}
    \label{rhoeq}
\end{equation}
where $T_C$ corresponds to the limiting temperature for the thermodynamics of the hadronic gas. For the case of $a>{7\over 2}$ (which is a case for models that account for the final size of hadrons e.g. Ref.\cite{Hagedorn:1980kb}),  within the simplified bootstrap model one can calculate the energy density of the phase transition to a quark-gluon system\cite{Hagedorn:1984hz}:
\begin{equation}
{\cal E}(T_C) = const \left ({kT_C\over 2\pi}\right) ^{3\over 2}
\left[{{3\over 2}kT_C\over (a-{5\over 2})}{1\over M_X^{a-{5\over 2}}}  + {1\over a-{7\over 2}}{1\over M_X^{a-7\over 2}}\right],
\label{Edenseq}
\end{equation}
where $const$ is the same in both Eq.(\ref{rhoeq}) and (\ref{Edenseq}).
In this model, the energy density at the phase transition is defined by $M_X$, and  it defines whether the transition is to the phase of a weakly interacting quark-gluon plasma (large ${\cal E}$) or to the  strongly interacting quark-gluon phase. Note that this phase cannot be a hadronic phase since it is evaluated at the maximum temperature at which hadronic system can not exist.

Even though the above discussion is confined strictly to the bootstrap model 
of the phase transition, it may stimulate the exploration of the connection between  parameters of CT and parameters that define a ``liberation" of  quarks in a high density and temperature hadronic system.

\subsection{Color Transparency and Vector Meson Dominance}
The vector meson dominance (VMD) model was a successful pre-QCD model in the description of diffractive photoproduction of vector mesons and electromagnetic interactions of hadrons in the multi-GeV energy regime. It was based on the observation that the physical photon can be expanded into the sum of a bare photon and a hadronic component\cite{Bauer:1977iq}:
\begin{equation}
\mid \gamma \rangle \approx \sqrt{Z_3}\mid \gamma_B\rangle + \sqrt{\alpha}\mid h\rangle
\end{equation}
in which contributions from the interaction of the bare photon are negligible for small angle diffractive scattering. The hadronic components should have the same quantum numbers of a photon $J^{PC} = 1^{--}, Q=B=S=0$ and thus, for the considered energies, such components are 
$\rho^0$, $\omega$ and $\phi$ mesons.

The VMD model emerges with the assumption that:
\begin{equation}
\sqrt{\alpha}\mid h\rangle = \sum\limits_{V=\rho,\omega,\phi} {e\over f_V}{m_V^2\over m_V^2 + Q^2}\mid V\rangle,
\label{VMDmodel}
\end{equation}
where $f_V$ is constant for each given vector meson related to the total width of the vector mesons' decay to an $e^-e^+$ pair. As mentioned by {Feynman\cite{Feynman:2019rot}}, this is a very bold  assumption, since there are several dynamical effects that should modify $f_V$ and imply a $Q^2$-dependence. These effects include the possibility of the photon coupling directly to $\pi$-mesons or various intermediate states, as well as effects of off-shellness of the vector mesons.

The post QCD situation for VMD models is not less bold since  within field theory descriptions of quark and gluon interactions, one expect photons to couple to $q\bar q$ states rather than the on-shell vector mesons. If such a coupling should proceed with the formation of real vector mesons during some time after the $\gamma\to q\bar q$ vertex, then one expects a formation process that can be similar to CT if the $q\bar q$ system at the origin is a PLC.

In the previous sections, we emphasized that CT effects require a formation of the PLC in which the relative transverse momentum of the valence quarks are on the order of ${1\over Q}$, enabling them to contain the gluon radiation between the minimal fock component quarks. Such effects question if heavy quark systems have a relative momentum $\sim {1\over m_q}$ at the production point. This is apparently the case for  $J/\Psi$-mesons whose photoproduction was investigated at 80-190~GeV energies at FermiLab\cite{FermilabTaggedPhotonSpectrometer:1986xzf}, and theoretical analysis indicated that it can be described by a photon converting to a $c\bar c$ pair before the target with the size of the $c\bar c$ pair smaller than the average $J/\Psi$ size. Formation of the $J/\Psi$ at these energies happens outside of the nucleus, and as a result the process, can be described by the $J/\Psi$-wave function at the origin\cite{Brodsky:1994kf}.

Potentially, one expects that a similar phenomenon could happen for $\phi$-mesons. At intermediate photon energies the formation of the $\phi$ mesons takes place in the nuclear medium- thus the measured rescattering cross section will result in a larger $\phi-N$ cross section  than one measured  at the  origin of $s\bar s$ production at the $\gamma \to s\bar s$ vertex.

Indeed, already in the 1970s it was observed that if one fixes the ${f_\phi^2\over {4\pi}}$ parameter from the $e^-e^+\to \phi$ process and evaluates the $\phi N$ cross sections within the description of VMD, one obtains a $\sigma_{\phi N}$ cross section of $9-12$~nb\cite{Bauer:1977iq, BEHREND197822,Sibirtsev2006}. However, if one  uses a quark model relating the $\phi N$ cross section to the combination of $K^{\pm}N$ and $\pi N$ cross sections\cite{Bauer:1977iq} where 
${f_\phi^2\over {4\pi}} \approx 13.2$ from storage ring measurements, then within VMD one obtains a twice larger cross section for the $\phi$ photoproduction.  The same discrepancy is observed if one estimates the $\phi N$ cross section from nuclear $\phi$-photoproduction data. 
A measurement studying the incoherent $\phi N$ interaction from heavier nuclei of Li, C, Al, Cu extracted a $\sigma_{\phi N}\approx 35$~mb~\cite{ISHIKAWA2005215} after accounting for the $A-$dependence of the $\phi$ photoproduction yield. The large increase in the $\sigma_{\phi N}$ as compared to the estimation for photoproduction from a single nucleon within VMD indicates that the $\phi$-mesons which reinteract in the nuclear medium are fully formed and have larger cross sections than the ``$\phi$" meson from the origin of $\gamma\to s\bar s$ transition.

Coherent $\phi$ production in deuterium observed large rescattering contributions for large values of $-t$~\cite{PhysRevC.76.052202} and was essentially consistent with previous coherent measurements on the proton extracting a $\sigma_{\phi N}=10$~mb assuming VMD. This study established that the $t-$slope may be more informative for extracting the fundamental $\sigma_{\phi N}$ cross section as a larger $t-$slope could also be consistent with the larger $\sigma_{\phi N}$ from the incoherent measurements. 

The discrepancy described above could be a manifestation of the CT phenomenon in which the $s\bar s$ state at the transition vertex is a PLC evolving to an on-shell $\phi$ meson which interacts in the nuclear medium. A future measurement of the  $\phi$-nucleon cross section  to solve the longstanding puzzle of whether the cross section is 10~mb, as extracted from photoproduction, or 30~mb, as obtained from nuclear rescattering, is presented in Ref.~\cite{Dalton:2025fbh}. A tensor polarized deuteron target would be used.

\subsection{Proton and heavy-ion collisions at RHIC and the LHC}
A more recent global analysis of proton or deuteron plus nucleus ($p/d+A$) collisions at both RHIC and the LHC studied configurations of the proton with a large-$x$ parton (i.e. $x_p\ge0.1$) that exhibited a smaller transverse size and subsequent reduced interactions with nuclear matter as quantified by a decrease of the number of nucleon-nucleon $NN$ interactions between the projectile and the target nucleus~\cite{PhysRevD.98.071502, PhysRevC.94.024915}. In collisions triggered on large-$x_p$ partons, the observed soft particle multiplicities and jet yields are consistent with a reduced interaction strength for small-size proton configurations. This effect was observed to grow with $x_p$, meaning that configurations carrying larger momentum fractions interact less strongly, consistent with color screening and CT expectations. The extracted interaction factor, ($\lambda(x_p)$, is the ratio of cross sections of large $x_p$ relative to the $NN$ cross section) and decreases with increasing $x_p$, showing that the proton becomes effectively more ``transparent" at larger $x_p$. At first glance, this discussion seems to be inconsistent with the analysis of Sect.~\ref{endpoint} explaining  that large values of $x_p$ corresponded to large sized configurations and a lack of CT. The apparent difference arises from the definition of the term ``large". In the Feynman mechanism large values of $x_p$ means $x_p=1-\Lambda_{\rm QCD}/Q$. Meaning, values of $x_p$ are very close to unity. In Ref.\cite{PhysRevD.98.071502}, values of $x_p$ range from about $0.1$ to 0.7, so that there is no contradiction. Indeed, the word `large' refers to large compared to values $\approx 10^{-5}$ that are important at high energies.

Interestingly, although small-sized configurations interact less, their interaction strength was observed to increase with collision energy (ascribed to the growing gluon densities at small $x_p$), but the suppression relative to average configurations remains substantial, especially at the lower energies explored. It was observed that configurations with a large-$x_p$ parton are naturally more color transparent, but their interaction strength for fixed $x_p$ is also reduced for lower energies which may be consistent with the energy regime for the observations of the EMC Effect~\cite{Geesaman:1995yd,Norton:2003cb,Hen:2013oha,Rith:2014tma,Malace:2014uea,Hen:2016kwk}. The region of $x_p$ between $\approx 0.3$ and 0.7 is where $\lambda(x_p)$ is smallest. This corresponds to the region of the EMC effect - the reduction of  quark distributions of bound nucleons  relative to that of free ones. One common explanation of the EMC effect \cite{Frankfurt:1985cv,PhysRevC.106.055202} is that configurations with nucleons of average size are enhanced in the nucleus due to attractive interactions. Probability conservation means that smaller sized configurations are suppressed. Such configurations correspond to fewer partons and therefore larger average values of $x_p$.

\subsection{Double scattering}
Exclusive processes in deuterium are well-described by the Generalized Eikonal Approximation(GEA)~\cite{BOEGLIN2024138742, Capel:2019zor, Sargsian:2004tz}. Due to the simplicity of the initial nuclear system being composed of only a proton and neutron, GEA can precisely describe the interaction between the struck nucleon and the spectator nucleon after the initial scattering reaction. 

For $Q^2>1$~GeV$^2$, the main amplitudes that describe the scattering process are the plane wave impulse approximation (PWIA) which dominates for initial nucleon momenta less than 200~MeV$/c$ and the re-scattering amplitude~\cite{BOEGLIN2024138742}. The PWIA contribution decreases at a faster rate than the re-scattering process with increasing initial nucleon momentum and thus, the re-scattering amplitude dominates for higher initial nucleon momenta. Double scattering is the squared re-scattering amplitude of this interaction between the nucleons and would be suppressed in the presence of PLCs. Double scattering is relevant for inter-nucleon distances of 1-2~fm.

 For small initial nucleon momentum, a deuterium target for traditional CT studies is not ideal due to the largely reduced contribution of FSIs. However, precisely selected kinematics for which the recoiling spectator nucleon angle is approximately perpendicular to the momentum transfer of the reaction and the initial nucleon momentum is $>300$~MeV/$c$, the double scattering contribution is enhanced beyond the PWIA contribution. In this way, the reduced inter-nucleon distances between the nucleon and its spectator lead to higher contributions of re-scattering~\cite{Frankfurt:1994kt}. Not only can these contributions be measured experimentally, but they would be minimized for PLCs in the presence of CT~\cite{physics4040092}. Such a measurement enables an observation of CT phenomena and also provides a method to evaluate their expansion rate which is expected to be rapid from recent experiments~\cite{Sargsian.2003}.

\section{Future experimental efforts to explore color transparency}\label{sect:future}
At this time, there are planned/ongoing experimental thrusts at Jefferson Lab that explore CT phenomena: 1) CT effects in mesons are studied by extending the $Q^2$ range for $\rho^0$ electroproduction in CLAS12 and $\pi^+$ electroproduction in the Hall C spectrometers and 2) CT effects in protons in rescattering kinematics will be studied in the Hall C spectrometers and 3) the process-dependence of CT is studied in photoproduction using the coherent bremsstrahlung beam in Hall D on various nuclei. 

\subsection{Meson electroproduction}
In the mesonic sector, confirmation of the continued increase in the $\rho^0$ and $\pi^+$ transparencies will be crucial for the interpretation of the expansion rate of PLCs. Recently, the CLAS Collaboration collected data with the CLAS12 spectrometer at Jefferson Lab using the Hall B cryogenic target and nuclear-foil assemblies to study $\rho^0$ electroproduction.  
\begin{figure}[htpb]
    \vspace{-15pt}
\centering
\includegraphics[width=13cm]{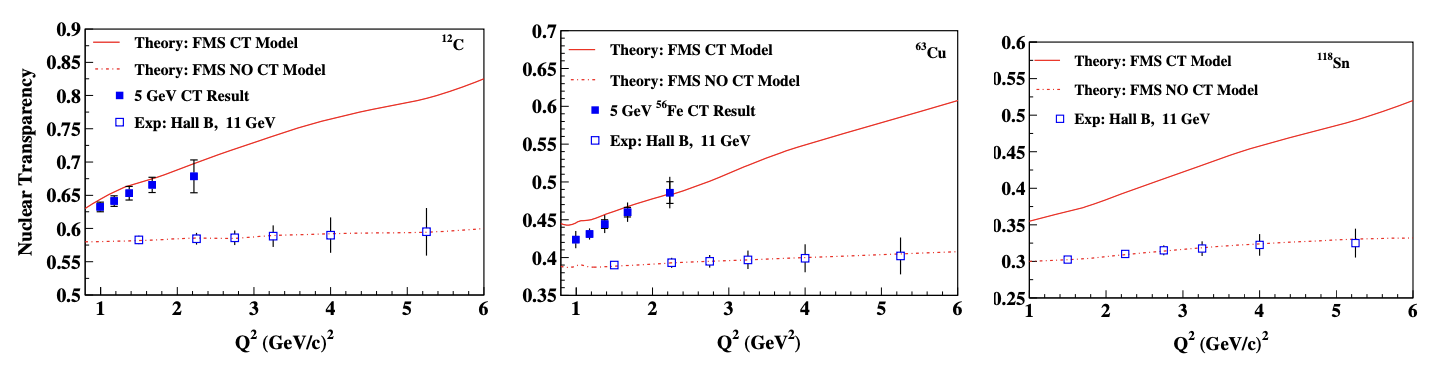}
\caption{ The CT projections for $^{12}$C (right), $^{63}$Cu (middle), and $^{118}$Sn (left) for lowest $l_c$ bin. Also included are the 5 GeV $^{12}$C and $^{56}$Fe CT results~\cite{ElFassi:2012nr} along with the FMS~\cite{Frankfurt:2008pz} model predictions tailored to the kinematics of the experiment. The model includes CT effects, FSIs, and $\rho^0$ decay and reproduces the 5~GeV CT results using a $\Delta M^2=$ 0.7 GeV$^2$.}\label{CT11GeV}
\end{figure}
This experiment extended the study of the CT phenomenon in the exclusive diffractive $\rho^0$ electroproduction off nuclei: $^{12}$C, $^{63}$Cu and $^{120}$Sn. The kinematics of the experiment were chosen to ensure small and fixed coherence length ($l_c$) of the $\rho^0$, in order to avoid the well known $l_c$ dependence of the nuclear transparency. This experiment extended the Q$^2$ range to much higher values, allowing a significant increase in the momentum and energy transfer involved in the reaction. Therefore, it is expected to produce much smaller configurations that live longer, expand slower, and exit the medium intact - the three primary pillars for CT studies. In addition, the measurements on several nuclei with different sizes will allow studying the space-time properties of the small size configuration (SSC) during its evolution to a full-sized hadron. The projected results are shown in Fig.~\ref{CT11GeV}.

\begin{figure}[htpb]
\begin{center}
\includegraphics[width=2.in]{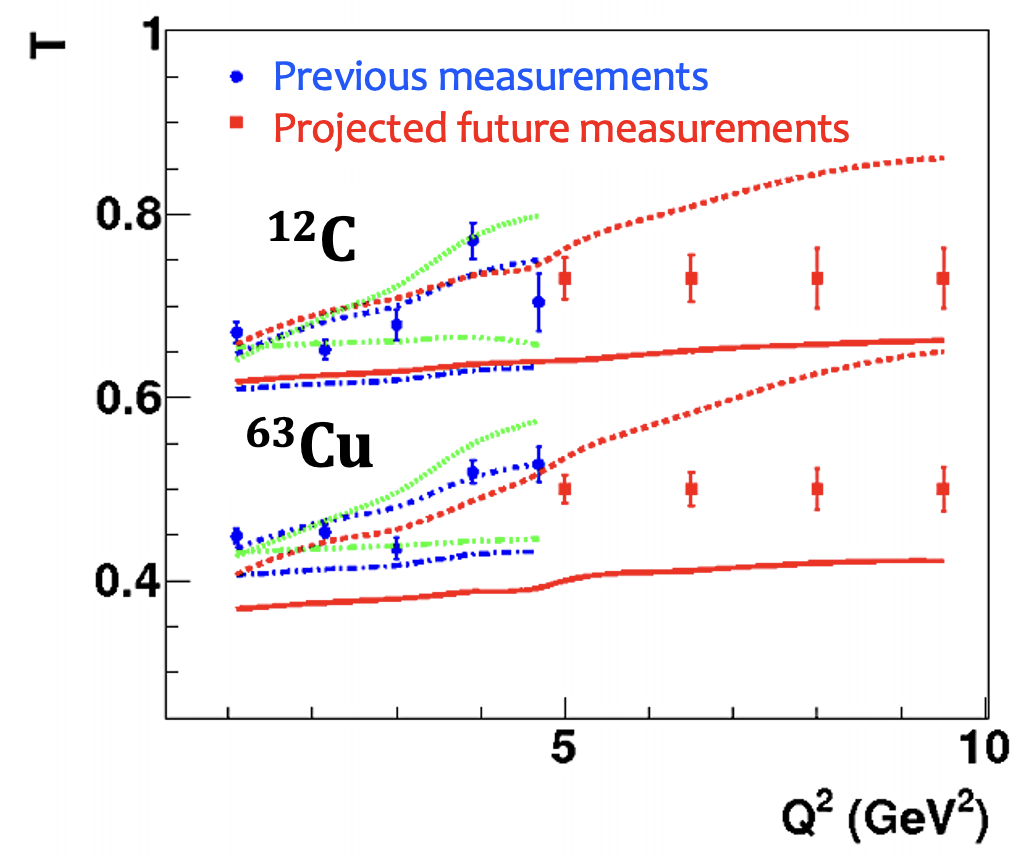}\includegraphics[width=1.9in]{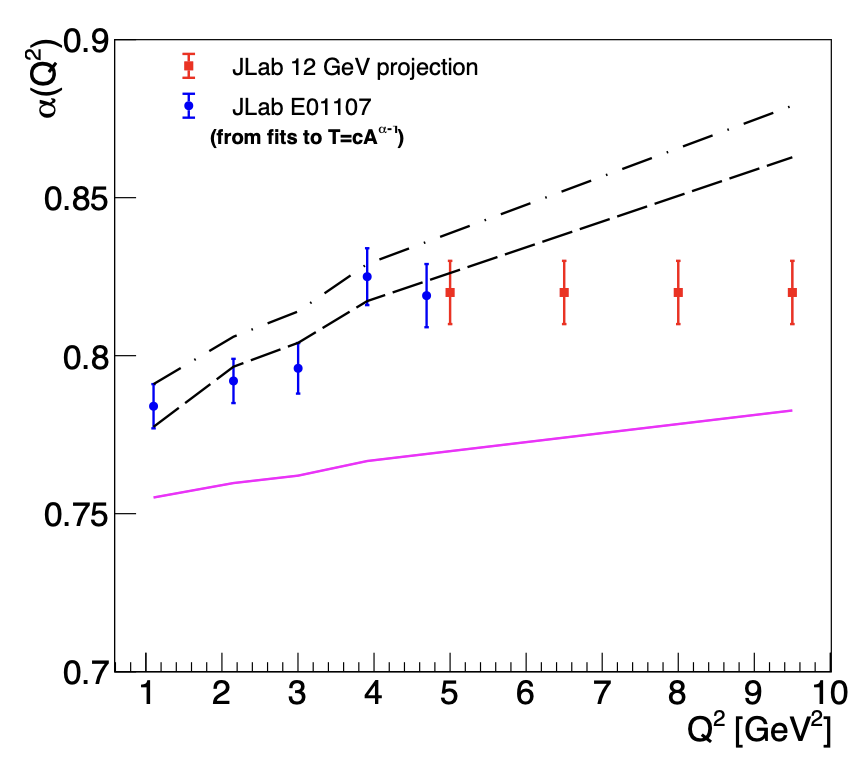}
\caption[]{\label{fig:pion-jlab} \baselineskip 13pt The previous measurements of the pion electroproduction transparencies are shown in blue and the projected future kinematic points are shown in red. Left: The transparencies are shown for carbon and copper targets as a function of $Q^2$ assuming a $\Delta M^2=$ 0.7 GeV$^2$. Right: The nuclear dependence, $A$, of the transparencies is shown as extracted from $T=cA^{\alpha-1}$ as a function of $Q^2$.}
\end{center}
\end{figure}

Likewise, the upcoming scheduled experiment in Hall C at Jefferson Lab for the $\pi^+$ will be able to access the highest $Q^2=9$~GeV$^2$ which may be able to connect the observation of CT to the shape and characteristics of the pion form factor. The kinematics that extend the previous measurement in the future experiment are shown in Fig.~\ref{fig:pion-jlab}. 
The pion electroproduction reaction mechanism may observe separate energy or $A$-dependencies which would be unrelated to CT. However, only CT predicts an increase in the transparency both with increasing energy and $A$.
A saturation of the transparency is also predicted, but the exact $Q^2$ for this observation is not yet known. Consequently, it is critical to measure the change in transparency with $Q^2$, as well as its nuclear $A$ dependence. Therefore, this experiment will measure pion-electroproduction from $^{12}$C and $^{63}$Cu targets in the Hall C spectrometers. The kinematics are  restricted to $|t|<1$~GeV$^2$ where the FSIs are reduced.

\subsection{Protons in rescattering kinematics}
All previous null observations of CT in the proton from quasielastic scattering utilized kinematics where the knocked-out proton is parallel to the $\vec{q}$ and already in a regime of minimal final state interactions. In such experiments, the observation of CT would be observed as an increase in the transmission due to the reduced absorption. A future experiment is planned to explore proton CT in maximal rescattering kinematics from a deuterium target which would effectively enhance the signal for the observation of CT. The benefit of such a reaction is that the production of the point-like configuration can be studied somewhat separately from its subsequent expansion by controlling the inter-nucleon distances after production~\cite{physics4040092}. 

\begin{figure}[htpb]
\begin{center}
\includegraphics[width=0.35\textwidth]{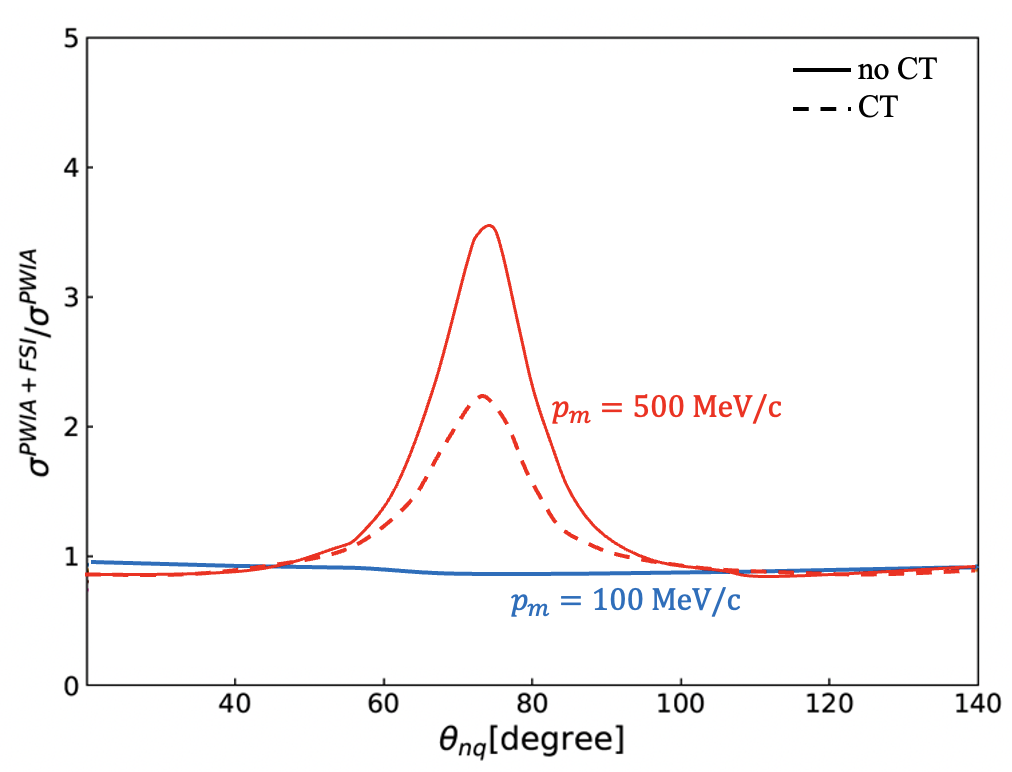}\includegraphics[width=3.3in]{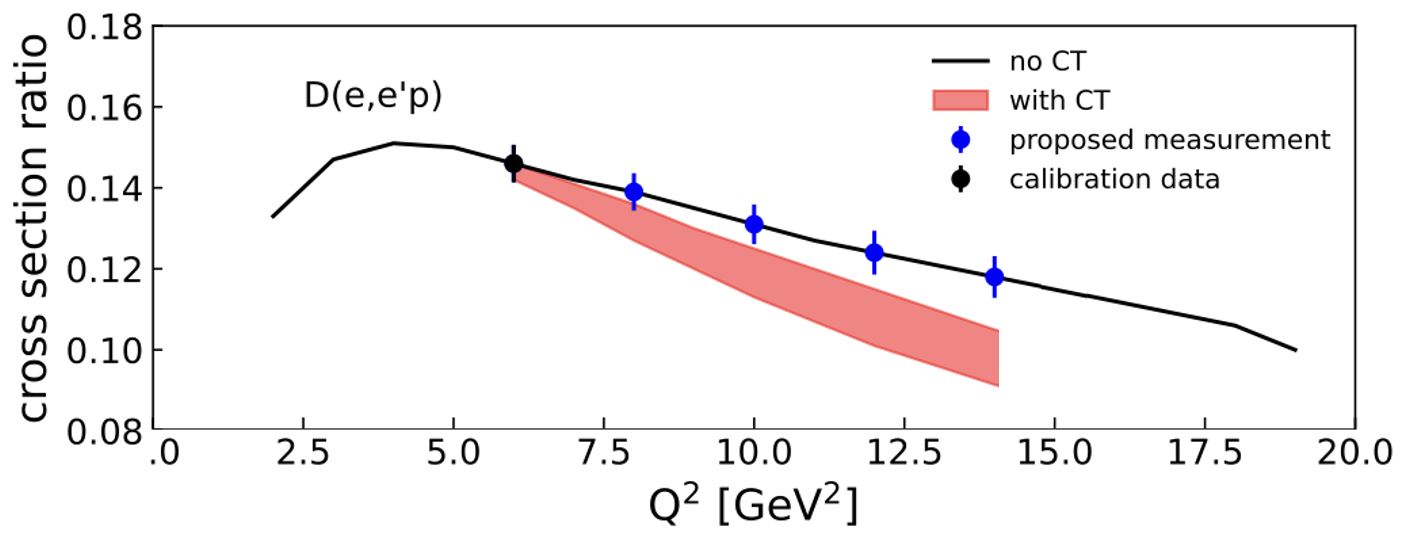}
\caption[]{\label{fig:ct-resc-exp} Left: For large $P_{miss}$ as shown, a signal for CT would appear as a reduction (dashed line) as compared to the nominal (solid line) in expected events from the rescattering peak. CT effects are larger for higher $P_{miss}$. Right: A future Jefferson Lab Hall C experiment will look for the onset of CT in protons by measuring a reduction in the ratio of protons from high $P_{miss}$ where rescattering effects are expected to dominate as compared to low $P_{miss}$.}
\end{center}
\end{figure}

Rescattering kinematics can effectively enhance the signal for the observation of CT by searching in regimes where FSI contributions are large and then observing an enhanced reduction in the FSIs under CT conditions. This effect is more significant in rescattering kinematics than in traditional experimental kinematics where the proton is knocked out along the $\vec{q}$ (parallel kinematics) with minimal FSIs. Assuming a $\Delta M^2=2$~GeV$^2$, one could observe a reduction in the measured rescattering events as shown on the left panel of Fig.~\ref{fig:ct-resc-exp}. Here, $\theta_{nq}$ indicates the angle between the reconstructed (undetected) neutron and the $\vec{q}$. The missing momentum, $\vec P_{miss}=|\vec P-\vec q|$ where $\vec P$ is the measured proton's momentum and the $\vec q$ is the difference in momentum between the incidenct and scattered electron.
Rescattering contributions are minimal for small $P_{miss}<200$~MeV/$c$. Therefore, the ratio of events measured at the maximal rescattering peak for high $P_{miss}$ to low $P_{miss}$ is a sensitive quantity to search for an overall reduction of FSIs with increasing $Q^2$. As shown on the right panel of Fig.~\ref{fig:ct-resc-exp}, a future experiment~\cite{physics4040092} at Jefferson Lab will look for the signal for the onset of CT in Hall C at Jefferson Lab in an overlapping $Q^2$ regime as~\cite{PhysRevLett.126.082301}, but with an enhanced sensitivity for larger $\Delta M^2$, hence accessing shorter PLC lifetime and more rapid expansion. 

\subsection{Photoproduction processes}

In photoproduction processes, the photon may interact as a ``resolved" photon (i.e. vector-meson) or as a direct photon (i.e. unresolved or point-like). At sufficiently high enough momentum transfers, $t$, it is expected that the direct photon interaction dominates the process and is able to uniformly sample throughout the nuclear volume~\cite{Larionov:2016mim}. The outgoing process may be described by Glauber multiple scattering or postulates that the outgoing meson is somewhat squeezed and experiences reduced interactions via CT effects. In the CT prediction, the reaction vertices can be produced from any point within the nuclear volume and will not re-interact while emerging form the nucleus, inducing an increase of the measured yield.

While most of the experimental searches for observing PLCs have been conducted in lepton and hadron hard scattering processes, the photonuclear process transfers the entire energy of the photon to the nucleon in the reaction, thus ensuring the highest achievable ``freezing" of the PLC in the reaction.

Hall D at Jefferson Lab produces a photon beam from coherent bremsstrahlung on a diamond radiator. The Hall D photon energies vary from about 8--11~GeV/$c$, sampling a unique phase space as compared to the other Jefferson Lab electroproduction experiments.

A recent experiment ran in Hall D at Jefferson Lab, measuring photonuclear
\begin{wrapfigure}[18]{r}{0.45\textwidth}
  \vspace{-18pt}
    \includegraphics[width=0.4\textwidth]{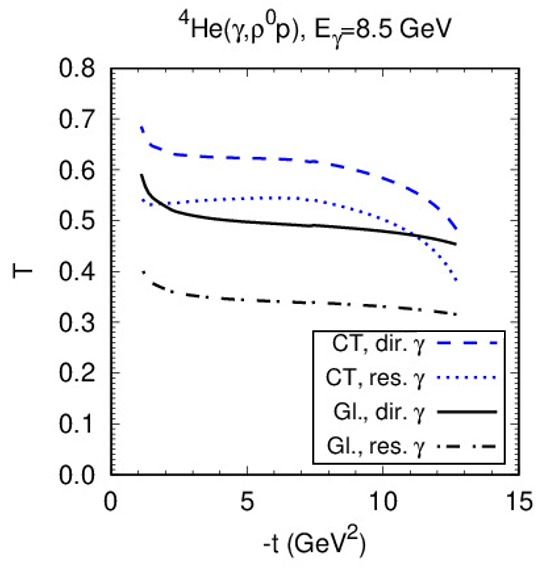}
  \vspace{-10pt}
  \caption{\label{fig:photo-rho} Transparency calculations~\cite{larionovEmail,Larionov:2016mim} for $^4$He$(\gamma,\rho^0 p)$ using the current Hall D photon beam  energy where $|u|>1$~GeV$^2$ are shown. The ``direct" and ``resolved" photon describe how the photon interacts in the hard process, while the Glauber and CT designations describe the final state treatment.}
\end{wrapfigure}
 reactions on deuterium, $^4$He, and $^{12}$C enabling access to numerous reaction channels that are currently under analysis. One such channel is the $^4$He($\gamma p,\rho^0 p$). Preliminary studies within particle identification and statistical constraints indicates that the transparency can be extracted up to $|t|\le 5$~GeV$^2$. The recent theory calculations from \cite{Larionov:2016mim} for $^4$He($\gamma p,\rho^0 p$) in these kinematics indicate that not only do CT effects play a role in the interpretation of the transparency but that the direct vs resolved interaction of the initial photon is also of interest (see Fig~\ref{fig:photo-rho}). The resolved photon calculation shown in Fig.~\ref{fig:photo-rho} assumes $\sigma_{\rho N}=25$~mb for the hard collision with the nucleon. The analysis of the data will yield a baseline for the photon transparency as well as the nuclear transparency of photo-produced $\rho^0$ mesons. Furthermore, this type of experiment can be extended to higher $t$ after the high-luminosity upgrade of the GlueX apparatus. 

\subsection{Prospects at J-PARC}
Hadronic collisions offer a complementary probe for exploring CT phenomena. The Japan Proton Accelerator Research Complex (J-PARC) has a high-intensity 30~GeV primary proton beams and 20~GeV/$c$ secondary pion beams. These beams make J-PARC uniquely suited to explore the intermediate energy regime where the onset of CT is expected to emerge and where previous results remain ambiguous.

One compelling motivation for CT studies at J-PARC lies in the opportunity to revisit  measurements from the BNL program, which observed an increase and subsequent unexpected decrease in nuclear transparency in $pA\rightarrow 2p(A-1)$ quasi-elastic scattering near 10~GeV/$c$~\cite{Leksanov:2001ui,Aclander:2004zm}. J-PARC can test whether these puzzling features persist and identify the conditions under which CT sets in. With a 30~GeV proton beam, J-PARC can probe the higher effective proton beam momenta needed to clarify the origins of the decrease in the proton transparency at high momenta~\cite{physics4020038,physics4020037}. The 30~GeV beam will also help extend the $(p,2p)$ measurements to higher energies beyond that obtained by the BNL experiments. This would test the idea that the nucleus filters out oscillations in the energy dependence of the $p-p$ cross section leading to oscillations in the nuclear transparency that are the inverse of the oscillations recorded in $p-p$ scattering experiments~\cite{Ralston:1990jj}. The oscillations in the $p-p$ scattering data were measured up to energies of $\sim$ 20~GeV, while the previous BNL $(p,2p)$ experiments could only reach $\sim$ 12~GeV. By extending these measurements to 20~GeV, the nuclear filtering idea can be tested rigorously.

A fixed-target proton beam experiment at J-PARC could also access new sensitive observables in the $p+D \rightarrow p+p+n$ reaction, where the two protons are fast while the neutron is a slow moving spectator. It has been shown that the ratio of cross sections with different proton transverse momenta and the azimuthal dependence of the cross section for this process is very sensitive to CT~\cite{Frankfurt:1997b}. These observables could be measured using the proton beam at J-PARC incident on a deuteron target. Both of the above mentioned experiments with proton beams would require a new high resolution spectrometer for the fast protons and a pion veto detector.

In addition to traditional two-body reactions, J-PARC is well-positioned to explore more advanced two-to-three hard exclusive processes, such as $\pi^-A\rightarrow\pi^-\pi^+A^*$, which offer a handle on the transverse size and evolution of hadronic configurations~\cite{KUMANO2010259}. In this approach, two high-momentum hadrons are produced alongside a low-momentum nuclear remnant, which allows for a better separation of the hard scattering subprocess from the nuclear effects. This method enables flexibility in tuning variables such as the coherence length and the transverse momentum which is extremely useful for probing the onset of CT across different nuclei and energy scales. These processes allow for an independent test of the CT mechanism by controlling the space-time evolution of the compact states and can decouple hadron formation time from expansion effects, making them a useful diagnostic for CT~\cite{physics4020037}. These kinematics would complement the approach to exploring the onset of CT using electron beams in rescattering kinematics at JLab~\cite{physics4040092}.

J-PARC's beam energies fill the gap between the lower-energy electron facilities and the higher-energy hadron machines such as COMPASS or the planned AMBER program at CERN. The ability to test CT using multiple beam species (protons and pions), with varied kinematic coverage and a strong experimental infrastructure, positions J-PARC to study the outstanding questions about the onset and nature of CT in QCD. Currently, there are no proposed nor scheduled experiments to directly measure CT at J-PARC. 

A recent workshop organized by the Center for Frontiers in Nuclear Science at Stony Brook University~\cite{agsworkshop} explored the possibility of using the BNL AGS to conduct fixed-target proton-nucleus collisions at intermediate energies. Such a facility would enable access to proton beam momenta of up to 24~GeV$/c$ (overlapping with proton beam energies available from J-PARC) and could extend the earlier BNL $(p,pp)$ measurements~\cite{Mardor:1998zf,Leksanov:2001ui,Aclander:2004zm}. As with the possible program at J-PARC, it would rigorously test the concept of nuclear filtering and help definitively interpret the trends in the transparency observed by the previous BNL experiments. However, it should be noted that the J-PARC facility is already operational while the BNL beamline is still at a conceptual stage. To mount an experiment at either J-PARC or the AGS, a new high resolution spectrometer would have to be built  with capabilities to detect fast protons and veto fast pions.

\subsection{Opportunities at future facilities}
We previously discussed upcoming experiments or not-yet published results from ongoing experiments including prospectives with current facilities. Here we comment on potential CT experiments that could be performed at possible and planned future facilities. 

\subsubsection{JLab at 22~GeV}
Current CT experiments will motivate the kinematics and requirements for experiments that can be accomplished if JLab is upgraded with a 22~GeV electron beam~\cite{accardi2023stronginteractionphysicsluminosity}. In the mesonic sector for exploring CT effects, confirmation of the continued rise in transparencies of the $\rho^0$ and $\pi^+$ using the 12~GeV electron beam in current experiments will need to be fully evaluated to further motivate studies at 22~GeV.

With an upgraded beam energy and the same targets and kinematics as used in the 12~GeV running of CLAS Run Group D, it is kinematically accessible to extend the $Q^2$ transparency dependence of the $\rho^0$ in Hall B up to 14~GeV$^2$. In Hall C, it is possible to extend the $Q^2$ dependence of the $A(e,e'\pi^+)n$ measurements up to 12.5~GeV$^2$ where the restriction in kinematics is in keeping $t<1$~GeV$^2$ in the current spectrometers to maintain a state of minimal FSIs.

 With a higher beam energy and the standard Hall C spectrometers, the $Q^2$ in the kinematics from the previous experiment~\cite{PhysRevLett.126.082301} can be extended up to about 17.4~GeV$^2$ on a carbon target. This increase in $Q^2$ can probe the CT hypothesis for more rapid expansions of the PLC proton and manifested higher $Q^2$ onset. Furthermore, in the near-term a 12~GeV experiment will explore proton CT in deuteron rescattering kinematics~\cite{physics4040092}. With a JLab 22~GeV beam upgrade, such an experiment would be able to extend the maximum $Q^2$ attainable to as high as 17~GeV$^2$ with no upgrades to the Hall C spectrometers. The need to attain these higher $Q^2$ transparency measurements for the proton will be better constrained from the results of the meson and proton studies in the current JLab 12~GeV program.

A 22~GeV beam energy upgrade at JLab would also facilitate higher energy studies of CT in photonuclear reactions. The recent Hall D experiment E12-19-003 studied photoproduction on nuclear targets of $^{12}$C, $^4$He, and deuterium. The results of this experiment observed sub-threshold $J/\Psi$ photoproduction on nuclear targets~\cite{PhysRevLett.134.201903}. The CT photoproduction analysis to look for PLC effects in the reaction of $\gamma~p\rightarrow\rho^0 p$ is currently in progress.

While the particle identification in the GlueX detector and statistics is limited to a $|t|$ range up to about 5~GeV$^2$ using the current setup, the data will be sensitive enough to distinguish between CT and non-CT effects as well as photon interactions as point-like and vector meson. Further evaluation is needed to optimize the coherent photon peak for reactions at 22~GeV and to understand the particle identification limitations or needs for improvements. Modest assumptions that would easily access a coherent photon peak around 15.5~GeV (as compared to the current 8.5~GeV) show that for the same reaction $^4$He$(\gamma p,\rho^0p)$ in Fig.~\ref{fig:photo-rho}, the accessible range of larger $|t|$ could easily go as high as 20~GeV$^2$ for $|u|>1$~GeV$^2$.  

Additional reaction channels remain to be studied in the current analysis, and their results could further motivate studies at the 22~GeV upgrade. Furthermore, at 22~GeV, $J/\Psi$ is accessible in both electro- and photoproduction reactions, which could provide another channel to study in the mesonic sector as well as provide a cross-check between different reaction mechanisms. 

\subsubsection{AMBER}
AMBER (Apparatus for Meson and Baryon Experimental Research) is the next generation experiment after COMPASS to use CERN's Super Proton Synchrotron (SPS) beamline\cite{amber_prop}. AMBER's beamline consists of a 400~GeV proton beam on a fixed target that generates high intensity secondary beams of muons, protons, pions and kaons in the ranges of 50-280~GeV/$c$. The AMBER spectrometer could be used with the proton beam on the liquid $^4$He target to study the $^4$He$(p,2p)$ reaction over the 50 - 280 GeV range thus extending the old BNL (p,2p) measurements to significantly higher energies. The liquid $^4$He target and spectrometer are already being built to measure the  anti-proton production cross sections. The same apparatus could be used to measure two outgoing protons in coincidence. Such an experiment could help identify the energy threshold for the onset of CT in protons.  

Further, AMBER is well-suited to  study the pionic PLC in quasielastic proton knockout from a nucleus using incident high-energy pions~\cite{PhysRevC.82.025205}. This process would be able to uniquely observe the pion as a PLC and would be able to study the cross section of the reaction as a function of $t$. At the energies available in AMBER, the PLC is not anticipated to expand while traversing the nucleus and so is studies are not focused on the onset of CT phenomena. Calculations for the transparency of $^{208}$Pb$(\pi,\pi p)$ relative to PWIA anticipate an enormous change in the transparency in a range of $|t|<10$~GeV$^2$~\cite{PhysRevC.82.025205}. 

The muon beam at AMBER~\cite{amber_prop} will be used with an active target time projection chamber (TPC) to measure the proton charge radius by detecting the recoiling proton. The same TPC could be used with an inert gas active target such as $^4$He, $^{20}$Ne or $^{40}$Ar to measure the quasielastic knockout of protons by detecting the recoiling $^3$H, $^{19}$F or $^{39}$Y nuclei. Such an experiment would be complementary to the electron scattering experiments at JLab and would be able to reach much higher momentum transfers. These experiments would also have the potential to identify the energy threshold for the onset of CT in protons. The previous CT review~\cite{DUTTA20131} explores other CT phenomena at COMPASS that can be explored with AMBER.

\section{Conclusions}
Color transparency is a fundamental prediction of QCD, in which hadrons produced in high-momentum exclusive processes can traverse nuclear matter with reduced interactions due to their compact, color-neutral configuration. Over recent decades, experimental efforts have searched for direct evidence of the onset of this phenomenon through measurements of nuclear transparency in electroproduction, photoproduction, and proton collision processes. 

While recent searches in protons ruled out the observed onset of CT in quasielastic scattering reactions at large $Q^2$, insights from this experiment indicate that protons in a PLC may have a far shorter lifetime than previously thought or they may contribute far less to the form factors. New experiments are planned to explore proton CT in rescattering kinematics. Pion and rho-meson electroproduction experiments have shown promising signatures consistent with the onset of CT. Recent and upcoming experiments from Jefferson Lab at 12~GeV will extend the previous $\rho^0$ and $\pi^+$ studies to higher $Q^2$ and can explore more details about how CT works. 

In reviewing the current experimental landscape, we've highlighted the theoretical description of PLCs and outlined future and other facility opportunities to explore CT. Studies of CT at facilities such as AMBER, J-PARC and Jefferson Lab at 22~GeV would be able to explore CT effects in new beams and processes. Together, these efforts offer a path toward confirming color transparency as a manifestation of QCD in nuclei.

\section{Acknowledgments}
The authors would like to express gratitude for the useful conversations with Yong Zhao, ANL,and Keh-Fei Liu, U. of Kentucky \& LBL. This publication was supported by the Guido~Altarelli Award from the DIS2024 conference in Grenoble, France. This work was funded in part by the U.S. Department of Energy, including DOE grant DE-SC0022007, DEFG02-01ER41172 and contract AC05-06OR23177 under which Jefferson Science Associates, LLC operates Thomas Jefferson National Accelerator Facility.
MS thanks Gyumri Technology Center, Armenia where part of the work was completed.
\bibliographystyle{ws-ijmpaHSV}
\bibliography{biblio}{}

\end{document}